\newcommand{\gpm}[3]{$#1^{+#2}_{-#3}$}

\def\lsim{\mathrel{\hbox{\rlap{\lower.55ex \hbox {$\sim$}}\kern-.0em
\raise.4ex \hbox{$<$}}}} 
\def\gsim{\mathrel{\hbox{\rlap{\lower.55ex \hbox {$\sim$}}\kern-.0em
\raise.4ex \hbox{$>$}}}} 

\def\aj{{AJ}}                   % Astronomical Journal
\def\araa{{ARA\&A}}             % Annual Review of Astron and Astrophys
\def\apj{{ApJ}}                 % Astrophysical Journal
\def\apjl{{ApJ}}                % Astrophysical Journal, Letters
\def\apjs{{ApJS}}               % Astrophysical Journal, Supplement
\def\ao{{Appl.~Opt.}}           % Applied Optics
\def\aap{{A\&A}}                % Astronomy and Astrophysics
\def\mnras{{MNRAS}}             % Monthly Notices of the RAS
\def\nat{{Nature}}              % Nature
\documentclass{newaa}
\usepackage{natbib}
\usepackage{graphicx}

\bibpunct{(}{)}{;}{a}{}{,}

\begin{document}
   \title{The star-formation rate in the host of GRB\,990712}

   \author{P. M. Vreeswijk \inst{1}
          \and
          R. P. Fender \inst{1}
          \and
          M. A. Garrett \inst{2}
          \and
          S. J. Tingay \inst{3}
          \and
          A. S. Fruchter \inst{4}
          \and
          L. Kaper \inst{1}
          }

   \institute{Astronomical Institute `Anton Pannekoek', University of
     Amsterdam \& Center for High Energy Astrophysics, Kruislaan 403,
     1098 SJ Amsterdam, The Netherlands
     \and
     Joint Institute for VLBI in Europe (JIVE), Postbus 2, 7990 AA,
     Dwingeloo, The Netherlands
     \and
     Australia Telescope National Facility, Paul Wild
     Observatory, Locked Bag 194, Narrabri, NSW 2390, Australia
     \and
     Space Telescope Science Institute, 3700 San Martin Drive,
     Baltimore, MD 21218, USA
     }  

   \date{Submitted August 8, 2001; accepted October 23, 2001}
   
   \abstract{We have observed the host galaxy of GRB\,990712 at 1.4
GHz with the Australia Telescope Compact Array, to obtain an estimate
of its total star-formation rate. We do not detect a source at the
position of the host. The 2$\sigma$ upper limit of 70 $\mu$Jy implies
that the total star-formation rate is lower than 100 M$_{\odot}$
yr$^{-1}$, using conservative values for the spectral index and
cosmological parameters. This upper limit is in stark contrast with
recent reports of radio/submillimeter-determined star-formation rates
of $\sim$ 500 M$_{\odot}$ yr$^{-1}$ for two other GRB host
galaxies. Our observations present the deepest radio-determined
star-formation rate limit on a GRB host galaxy yet, and show that also
from the unobscured radio point-of-view, not every GRB host galaxy is
a vigorous starburst.

    }

    \maketitle

%
%________________________________________________________________

\section{Introduction}
\label{sec:introduction}

The gamma-ray bursts (GRBs) for which afterglows have been observed so
far, i.e. bursts with a duration longer than roughly two seconds, can
be most adequately explained by the collapse of a rapidly rotating,
massive star. In this collapsar model
\citep{1993ApJ...405..273W,1998ApJ...494L..45P,1999ApJ...524..262M,2001ApJ...550..410M},
the GRB is produced in narrow cones along the rotational axis of the
collapsing progenitor, accompanied by an isotropic supernova explosion
of a type similar to SN\,1998bw \citep{1998Natur.395..670G}.  For a
few GRBs the presence of a supernova has been inferred from a bump in
the optical afterglow light curve at late times
\citep{1999Natur.401..453B,1999ApJ...521L.111R,2000ApJ...536..185G}.
Moreover, the locations of GRB afterglows coincide with the optical
extent of their host galaxies \citep[][ Fruchter et al., in
prep.]{bloomsubm}, suggesting that these long-duration GRBs are linked
with regions of star formation. The observed GRB location distribution
is not expected for the alternative binary neutron star merger model
(or a neutron star and a black hole)
\citep{1989Natur.340..126E,1992ApJ...395L..83N}, where the kick
velocities received from the two supernovae and the time it takes the
two compact objects to merge, would cause the GRB to occur kiloparsecs
away from the place of birth of the progenitor binary
\citep{1999MNRAS.305..763B}, in at least a few cases. These mergers,
however, are expected to be the progenitors of the category of
short-duration GRBs \citep{1999ApJ...526..152F}.

In case the gamma rays come from internal shocks, which is the
generally favoured model, a GRB can be observed both in the case of a
collapsar and of a merger, i.e. irrespective of its environment.
Afterglows are thought to be produced by the interaction of the
fireball ejecta with the environment \citep[the flux in the fireball
model is proportional to the square root of the density of the
circumburst medium, e.g.][]{1999ApJ...523..177W}.  If mergers would
also produce long-duration GRBs, we would have expected a fraction of
these to have no X-ray afterglow -- namely those which occur outside a
galaxy. However, nearly all attempts to detect an X-ray afterglow were
successful \citep[e.g.][]{2000hgrb.symp..375S}, which suggests that
they are not the result of mergers (assuming the internal shock
model). Note that the location argument in favour of the collapsar
model does not necessarily hold if the gamma rays are produced by
external shocks. In that case the gamma rays themselves are produced
by interaction with the circumburst medium, which may mean that all
GRBs (both long and short-durations bursts) and their afterglows that
occur in low-density environments are not detected.

If GRBs are intimately connected with the deaths of massive stars,
they are potential probes of star formation in the early universe. At
present, it is not known which type of galaxy produces the bulk of
star formation at high redshift: the numerous faint blue galaxies
\citep{1997ARA&A..35..389E}, or the ultra-luminous infrared or
starburst galaxies \citep{1996ARA&A..34..749S}. Determination of the
type of galaxy that gives birth to GRBs can provide important clues to
this outstanding issue.

Star-formation rates (SFRs) for several GRB host galaxies have been
estimated from optical nebular emission lines (e. g. [O{\sc ii}]) to
vary from 0.3 M$_{\odot}$ yr$^{-1}$ for the host of GRB\,970828
\citep{djorgovski970828} to 24 M$_{\odot}$ yr$^{-1}$ for the host of
GRB\,980703 \citep{1998ApJ...508L..17D}. These values are not yet
corrected for dust extinction, which is difficult to estimate and
which can be quite large. This causes considerable uncertainty in the
SFR values.  Recently, very high star-formation rates ($\sim$ 500
M$_{\odot}$ yr$^{-1}$), have been inferred for GRB\,980703 and
GRB\,010222, using radio \citep{berger} and submillimeter
\citep{frail222} measurements, respectively, which do not suffer from
dust extinction.  The question that arises is: do all GRB host
galaxies look like vigorous starburst galaxies when they are observed
at the unobscured radio and submillimeter wavelengths?

Due to its relative proximity ($z$ = 0.433), the host of GRB\,990712
is an excellent GRB host galaxy to study in detail. VLT spectra of the
host show that the galaxy is an H{\sc ii} galaxy (i.e. the spectral
emission lines are produced by H{\sc ii} regions that are being
ionized by O and B stars) and not a galaxy that is hosting an active
galactic nucleus (AGN). The [O{\sc ii}] emission star-formation rate
has been inferred to be SFR$_{\rm [OII]} = $\gpm{35}{178}{25}
M$_{\odot}$ yr$^{-1}$ \citep{2001ApJ...546..672V}. The large errorbars
are due to the uncertainty in the estimate of the optical extinction.

To obtain an independent estimate of the SFR in the host galaxy of
GRB\,990712, we performed 1.4 GHz (21 cm) observations with the {\it
Australia Telescope Compact Array} (ATCA) in March 2001. The radio
continuum flux of a normal galaxy (i.e. a galaxy that is not hosting
an AGN) is thought to be produced by synchrotron radiation from
electrons which are accelerated by supernova remnants, and free-free
emission from H{\sc ii} regions \citep{1992ARA&A..30..575C}.  The
radio continuum emission should therefore be well-correlated with very
recent star formation, which is strongly supported by the observed
far-infrared/radio correlation.  The obvious advantage of this method
over the optical emission-line measurements is that the radio flux is
unaffected by dust extinction, allowing an unobscured view of the
star-formation nature of the GRB host.

%
%________________________________________________________________

\begin{figure}
  \centering
  \includegraphics[width=8cm,angle=-90]{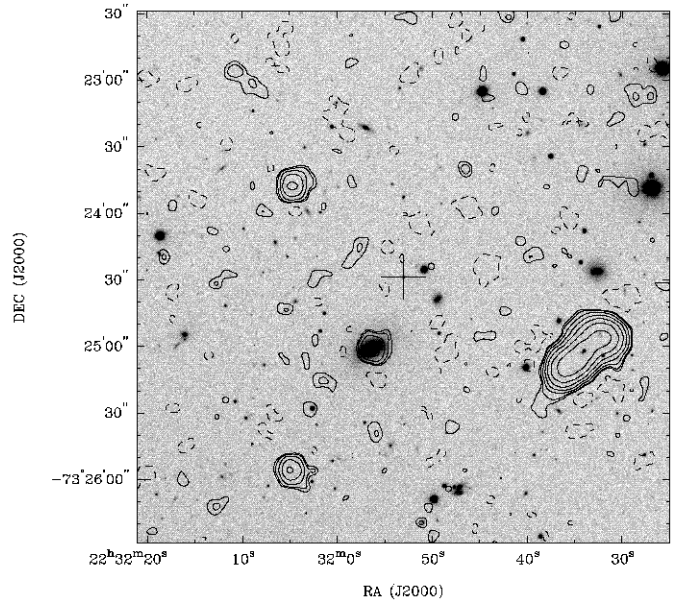}
  \caption{\small The field of the host galaxy of GRB\,990712 in
    the optical (greyscale) and radio (contours). The R-band image was
    taken with the VLT on 13 July 1999 \citep{2000ApJ...540...74S},
    only 0.5 days after the burst when the optical counterpart still
    outshone its host galaxy. Its position is indicated with a cross,
    The plotted radio contours are factors of $-$2 (dashed), 2, 3, 6,
    12, 24, 48, and 96 (all solid) times the noise level of 35 $\mu$Jy.
    No 1.4 GHz source is detected above 2$\sigma$ at the position of
    GRB\,990712.}
  \label{fig:radiomap}
\end{figure}

\section{Observations}
\label{sec:observations}

The host galaxy of GRB\,990712 was observed with the {\it Australia
Telescope Compact Array} (ATCA) between 2001 March 29, 15:06 (MJD
51997.63) and March 30 03:22 (MJD 51998.14). The observations were
performed in the 6D antenna configuration, in two bands centred at
1344 and 1432 MHz respectively, with a total observing time on-source
of 10.18 hr. Absolute flux calibration was achieved using PKS
1934-638; PKS 2101-715 was used as the phase calibrator. Data
reduction was performed using the MIRIAD software
\citep{1995adass...4..433S}. Figure 1 shows a uniformly-weighted map
of the region around the host galaxy with a beam size of
$7.7\times8.8$ arcsec, superposed on an optical image
\citep{2000ApJ...540...74S} taken with ESO's Very Large Telescope
(VLT) . No source is detected at the host galaxy location to a
2$\sigma$ limit of 70 $\mu$Jy -- the noise level of 35 $\mu$Jy was
estimated from measuring the sky around the target region. The
theoretical noise limit for the observing time, bandwidth, frequency
and array used is 20 $\mu$Jy/beam, i.e. less than a factor of two
lower than we obtain.  We were unable to achieve a lower noise level
using natural weighting because of sidelobes from bright, nearby
sources.

%
%________________________________________________________________

\section{Results}
\label{sec:results}

For a normal star-forming galaxy, the radio continuum emission is
proportional to the star-formation rate \citep{1992ARA&A..30..575C}:
\begin{equation}
\rm{SFR_{0.1-100 M_{\odot}}} = \frac{L_{\nu,\rm obs}{[\frac{1.4}{(1+z)
    \nu_{\rm GHz}}]}^{\alpha} \times 4\pi d_{\rm l}^2 \times 5.5}{5.3\times10^{28}\nu_{\rm GHz}^{\alpha} +5.5\times10^{27}\nu_{\rm GHz}^{-0.1}}
\end{equation}
where $L_{\nu,\rm obs}$ is the observed luminosity in $\mu$Jy at
$\nu_{\rm GHz}$ and the factor ${[\frac{1.4}{(1+z) \nu_{\rm
      GHz}}]}^{\alpha}$ converts this to the emitted luminosity at 1.4
GHz \citep[see][]{2000ApJ...544..641H}; $\alpha$ is the spectral index
(with S$_\nu \propto \nu^{\alpha}$), and $\rm d_l$ is the luminosity
distance in cm. The first term in the denominator represents the
dominant synchrotron component from electrons accelerated in supernova
shocks; the second term comes from the thermal emission from H{\sc ii}
regions \citep[for a derivation of both components,
see][]{1992ARA&A..30..575C}.  Strong evidence for this relation is
supplied by the correlation between the SFR determined through Eq.  1
and that obtained from the far-infrared luminosity \citep[see e.g.
Fig. 1 of][]{1998ApJ...507..155C}, which extends over four orders of
magnitude \citep[see also][ and references
therein]{1992ARA&A..30..575C}. The expected lifetime of the remnant
synchrotron emission is $\leq$ 10$^8$ years, and thus radio continuum
emission is a good tracer of very recent star formation. The original
Condon relation is valid for stars with masses above 5 M$_{\odot}$.
\citet{2000ApJ...544..641H} extend this to the range 0.1-100
M$_{\odot}$ by multiplying with a factor $Q$ = 5.5, which assumes a
Salpeter initial mass function (IMF), i.e.  $\psi(M) \propto
M^{-2.35}$. This factor becomes 5.3 if an upper mass cut-off of 125
M$_{\odot}$ is used, whereas it is 3.8 if the range of masses is
0.25-100 M$_{\odot}$. Changing the IMF slope with $\pm$ 0.1
\citep{1999ApJ...515..323E} results in the range $Q$ = 4-7.5.  For
starburst regions, where the IMF may be weighted toward high-mass
stars \citep{1999ApJ...515..323E}, $Q$ may be as low as unity.  We
take $Q$=5.5.

The resulting value for the star-formation rate that we obtain depends
on the assumed spectral index and the luminosity distance, hence on
the adopted cosmology. We therefore calculate the SFR upper limit with
the spectral index ranging from $-$0.35 to $-$1.0, typical for normal
radio galaxies, and the cosmologies ($h$, $\Omega_{\rm m}$,
$\Omega_{\lambda}$) = (0.65, 1, 0), (0.65, 0.2, 0), and (0.65, 0.3,
0.7), where $h$ is the dimensionless Hubble constant, defined as: $h$
= $\frac{\rm H_0}{\rm 100\,km\,s^{-1} Mpc^{-1}}$.  The redshift of the
GRB and its host galaxy is 0.4331 $\pm$ 0.0004
\citep{2001ApJ...546..672V}.  The resulting luminosity distances
\citep[see][]{hogg} for these cosmologies are respectively:
6.7$\times$10$^{27}$ cm, 7.3$\times$10$^{27}$ cm and
7.9$\times$10$^{27}$ cm.  Inserting our 2$\sigma$ upper limit of 70
$\mu$Jy, and the range in spectral indices and luminosity distances,
we obtain the following range for the upper limit on the SFR in the
host of GRB\,990712: SFR$_{\rm max}$ = 47-100 M$_{\odot}$ yr$^{-1}$.

\begin{table*}
  \caption[]{\small
    Star-formation rate estimates from the [OII] nebular emission-line and
    1.4 GHz flux; for both a Salpeter IMF over the mass range
    0.1-100 M$_{\odot}$ is assumed. The
    [OII] fluxes (corrected for the Galactic foreground 
    extinction), are taken from the following references:
    1. \citet{2001ApJ...554..678B}; 2. \citet{1998ApJ...507L..25B}; 3. 
    \citet{djorgovski970828}; 4. \citet{djorgovski613}; 
    5. \citet{1998ApJ...508L..17D}; 6. \citet{2001ApJ...546..672V}; 
    7. \citet{2001A&A...370..398C}; 8. \citet{price911}. We use
    the currently popular cosmology ($h$,
    $\Omega_{\rm m}$, $\Omega_{\lambda}$) = (0.65, 0.3, 0.7) to obtain the
    luminosity distance \citep[see also][]{2001AJ....121.2879B}, in order 
    to convert the line flux into luminosity. We apply the conversion 
    from \citet{1998ARA&A..36..189K} to this [OII] luminosity to obtain the 
    attenuated SFRs, listed in column 4. The error in this conversion
    is roughly 30\% \citep{1998ARA&A..36..189K}. We correct these
    SFRs in two different ways. First, we use the host extinction
    measurements from \citet{2001A&A...372..438S} and   
    \citet{2001ApJ...546..672V}, and scale the attenuated SFR to the
    extinction-corrected SFR of column 5. Second, the attenuated SFR 
    estimates are also corrected in a more general way, using the 
    prescription of \citet{2001AJ....122..288H} for the relation
    between the SFR obtained from optical emission lines, and the SFR
    obtained from the 1.4 GHz flux for a sample of radio
    galaxies. These estimates are shown in column 7. The 
    scatter in this conversion from attenuated to corrected SFR 
    \citep[see Fig. 3 of][]{2001AJ....122..288H} is approximately a
    factor of 2 to 3. The radio-determined SFRs are listed in column 6.}
\label{tab:sfr}
$$
\begin{array}{ccccccc}
    \hline
    \noalign{\smallskip}            
    \rm host\,of\,GRB & 
    \rm [OII]\,flux & 
    \rm reference & 
    \rm SFR_{\rm [OII]-att.} & 
    \rm SFR_{\rm [OII]-corr.} & 
    \rm SFR_{\rm 1.4GHz}^{\mathrm{d}} &
    \rm SFR_{\rm [OII]-Hopkins} \\ 
    & 
    {\rm (10^{-16} erg\,s^{-1} cm^{-2})} & 
    &
    \rm M_{\odot}\,yr^{-1} & 
    \rm M_{\odot}\,yr^{-1} & 
    \rm M_{\odot}\,yr^{-1} & 
    \rm M_{\odot}\,yr^{-1} \\
    \hline
    970228 & 0.22 \pm 0.01 & 1 & 0.76 & -                & < 380 ^{\mathrm{e}} & 5.5 \\
    970508 & 0.30 \pm 0.02 & 2 & 1.6  & 5.8^{\mathrm{b}} & < 430 ^{\mathrm{f}} & 17 \\
    970828 & 0.045 \pm 0.01^{\mathrm{a}} & 3 & 0.3        & -                & < 360 ^{\mathrm{g}} & 1.2 \\
    980613 & 0.44          & 4 & 4.7  & 21^{\mathrm{b}}  & - & 82 \\
    980703 & 3.04          & 5 & 24   & 34^{\mathrm{b}}  & \approx 500 & 784 \\
    990712 & 3.32 \pm 0.14 & 6 & 3.7  & 44^{\mathrm{c}}  & < 100  & 58 \\
    991208 & 1.79 \pm 0.22 & 7 & 6.4  & 23^{\mathrm{b}}  & - & 127 \\ 
    000911 & 0.23 \pm 0.03 & 8 & 2.2  & -                & - & 28 \\
    \noalign{\smallskip}
    \hline
  \end{array}
  $$
  \begin{list}{}{}
  \item[$^{\mathrm{a}}$] Assuming galaxy B of \citet{djorgovski970828}
    is the host. Inclusion of galaxy A would increase the [OII] flux
    and the attenuated SFR with a factor of about 5, resulting in a
    corrected SFR of 17 (column 7).
  \item[$^{\mathrm{b}}$] Adopting the host extinction as
    determined by \citet{2001A&A...372..438S}.
  \item[$^{\mathrm{c}}$] Adopting the host extinction as determined by
    \citet{2001ApJ...546..672V}, who used ($h$, $\Omega_{\rm m}$,
    $\Omega_{\lambda}$) = (0.70, 0.3, 0), which results in a lower SFR
    of 35 M$_{\odot}$ yr$^{-1}$.
  \item[$^{\mathrm{d}}$] All upper limits are 2$\sigma$, and we assume
    $\alpha$ = $-$1 and ($h$, $\Omega_{\rm m}$, $\Omega_{\lambda}$) =
    (0.65, 0.3, 0.7).
  \item[$^{\mathrm{e}}$] Taking the estimated 2$\sigma$ upper limit of 71
    $\mu$Jy from \citet{1998ApJ...502L.119F} and $z$=0.695.
  \item[$^{\mathrm{f}}$] Taking the estimated 2$\sigma$ upper limit of 45
    $\mu$Jy from \citet{2000ApJ...537..191F} and $z$=0.835.
  \item[$^{\mathrm{g}}$] Taking the estimated 2$\sigma$ upper limit of
    26 $\mu$Jy from \citet{djorgovski970828} and $z$=0.958.
  \end{list}
\end{table*}

%
%________________________________________________________________

\section{Discussion}
\label{sec:discussion}

Table 1 shows the GRB host galaxies for which an [O{\sc ii}]
emission-line flux has been reported in the literature. The H$\alpha$
line emission is a more reliable optical estimate of the SFR than
[O{\sc ii}], but this line is usually shifted into the near-infrared
passbands, making it observationally difficult to obtain the H$\alpha$
line flux. The listed [O{\sc ii}] flux, which is corrected for the
Galactic foreground extinction, is taken from the references indicated
in the caption. We convert this flux, using ($h$, $\Omega_{\rm m}$,
$\Omega_{\lambda}$) = (0.65, 0.3, 0.7) and the calibration from
\citet{1998ARA&A..36..189K}, to obtain an estimate of the attenuated
SFR (column 4), i. e. not corrected for dust extinction. The error in
this conversion is roughly 30\%. With ($h$, $\Omega_{\rm m}$,
$\Omega_{\lambda}$) = (0.65, 0.2, 0), these SFR estimates need to be
scaled down by about 20\%.

The [O{\sc ii}] emission line method shows that GRB host galaxies
differ widely in their star-formation nature, from 0.3 to 24
M$_{\odot}$ yr$^{-1}$, i.e. a factor of 80. However, extinction is an
important factor in optical SFR indicators, which could boost these
estimates up to much higher values. A good example is the host galaxy
featured in this paper: GRB\,990712.  \citet{2001ApJ...546..672V}
estimate the SFR of the host of GRB\,990712 to be \gpm{35}{178}{25}
M$_{\odot}$ yr$^{-1}$ from the [O{\sc ii}] emission line,
\gpm{64}{770}{54} M$_{\odot}$ yr$^{-1}$ from H$\beta$ and $\sim$ 400
M$_{\odot}$ yr$^{-1}$ from the 2800 \AA\ flux. These values reflect
the large uncertainty in the dust extinction estimate for this host,
obtained from the observed and expected ratio of H$\beta$ and
H$\gamma$. The radio data presented in this paper provide a clear
upper limit to these SFRs of 100 M$_{\odot}$ yr$^{-1}$, indicating
that the extinction at 2800 \AA\ has been overestimated. In Table
1 we also list the extinction-corrected SFRs, using
host-galaxy extinction measurements from the literature
\citep{2001A&A...372..438S,2001ApJ...546..672V}.  Note that the value
for the host of GRB\,990712 differs from that reported by
\citet{2001ApJ...546..672V} due to the different cosmologies used.

Recently, very high SFRs have been inferred for two other GRB host
galaxies, not from the optical but through two methods that are not
affected by the dust along the line of sight.  From millimeter and
submillimeter wavelength observations, \citet{frail222} infer a rate
of $\sim$ 600 M$_{\odot}$ yr$^{-1}$ for the host of GRB\,010222. The
host of this burst is faint: V = 26.0 $\pm$ 0.1
\citep{2001GCN..1087....1F}, making it difficult to obtain an [O{\sc
ii}] flux measurement.  \citet{berger} measure a 1.4 GHz flux of 68.0
$\pm$ 6.6 $\mu$Jy for the host of GRB\,980703 ($z$=0.966), and infer a
star-formation rate of $\sim$ 500 M$_{\odot}$ yr$^{-1}$, with the same
method and assumptions as we employ in this paper.  The latter authors
use the empirical relation between the far infra-red (FIR) and radio
emission \citep{1971A&A....15..110V,1985ApJ...298L...7H}, to estimate
the FIR luminosity of the host of GRB\,980703 to be that of an
ultra-luminous infra-red galaxy (ULIG or ULIRG, for which L$_{\rm IR}$
$>$ 10$^{12}$ L$_{\odot}$). Applying the same relation to our 1.4 GHz
flux of GRB\,990712 (using $\rm d_l$ = 7.9$\times$10$^{27}$ cm), we
find L$_{\rm IR} < 1.1\times$ 10$^{11}$ L$_{\odot}$, more than a
factor of 20 less luminous in the IR than the host of
GRB\,980703. This indicates that the host of GRB\,990712 does not
belong to the class of ULIRGs, although direct IR observations would
be needed to definitely rule out the possibility.

Our observations of the host of GRB\,990712 present the deepest
radio-determined SFR limit on a GRB host galaxy yet, and show that
also when observed at unobscured radio wavelengths, not every host is
a vigorous starburst galaxy. How do the [O{\sc ii}] and radio SFR
estimates compare in general? The optical SFR estimators, after
correcting for internal extinction, tend to underestimate the total
SFR, when compared to the FIR and radio methods \citep[e.g.
][]{1998ApJ...507..155C}. The host of GRB\,980703 is a striking
example of this: the extinction-corrected SFR is about a factor of 15
lower than the rate estimated from the radio continuum flux (see Table
1). We therefore use the prescription of \citet[][ Eq.
5]{2001AJ....122..288H} to convert the attenuated SFR estimates of
column 4 in Table 1 to total SFRs (column 7). This
conversion is based on an empirical correlation between obscuration
and far-infrared luminosity \citep{1996ApJ...457..645W}, from which
\citet{2001AJ....122..288H} deduce a relation between obscuration and
SFR. These numbers can now be compared with SFR methods that do not
suffer from dust extinction, such as the 1.4 GHz continuum flux
method. For the two host galaxies for which these numbers are
available, the values are consistent within the (large) errors.

A large sample of GRBs with redshift determinations (through e.g.
rapid spectroscopy of the afterglow) and 1.4 GHz or submillimeter
observations of their hosts, can provide an important step toward
calibration of the possible relation between GRB number counts and the
total star-formation density as a function of look-back time.
Different classes of GRBs, if produced by different progenitors, could
be used to verify this calibration.

\begin{acknowledgements}
  PMV is supported by the NWO Spinoza grant 08-0 to E.P.J. van den
  Heuvel. LK is supported by a fellowship of the Royal Academy of Arts
  \& Sciences in the Netherlands. The Australia Telescope Compact
  Array is part of the Australia Telescope which is funded by the
  Commonwealth of Australia for operation as a National Facility
  managed by CSIRO. PMV kindly thanks David Hogg for providing his
  code \citep[see][]{hogg} to verify the obtained luminosity
  distances. PMV also thanks E.P.J. van den Heuvel for valuable
  comments.
\end{acknowledgements}

\end{document}